\RequirePackage{fix-cm}
\documentclass[twocolumn]{svjour3}       
\usepackage{flushend} 
\smartqed  
\usepackage{amsmath}
\usepackage{graphicx}
\usepackage{xcolor}
\usepackage{mathrsfs}
\usepackage{txfonts}

\usepackage{hyperref}

\begin{document}

\title{Paving Through the Vacuum
}


\author{C.M.F. Hugon         \and
        V. Kulikovskiy 
}


\institute{C.M.F. Hugon \at
              R\&DoM, 1 avenue du Corail, 13008 Marseille, France \\
              \email{chr.hugon@protonmail.com}           
           \and
           V. Kulikovskiy \at
           INFN - Sezione di Genova, Via Dodecaneso 33, 16146 Genova, Italy
}


\maketitle

\begin{abstract}
 Using the model in which the vacuum is filled with virtual fermion pairs, we propose an effective description of photon propagation compatible with the wave-particle duality and the quantum field theory.

 In this model the origin of the vacuum permittivity and permeability appear naturally in the statistical description of the gas of the virtual pairs. Assuming virtual gas in thermal equilibrium at temperature corresponding to the Higgs field vacuum expectation value, $kT\approx246.22\, \rm{GeV}$, the deduced value of the vacuum magnetic permeability (magnetic constant) appears to be of the same order as the experimental value. One of the features that makes this model attractive is the expected fluctuation of the speed of light propagation that is  at the level of $\sigma\approx 1.9\,\mathrm{as\,m}^{-1/2}$. This non-classical light propagation property is reachable with the available technologies.

\end{abstract}

\section{Introduction}
\label{s:introduction}
Following \cite{RefUrban}, we develop a new model in which the virtual fermions are continuously appearing and disappearing by pairs of particle-antiparticle.

The virtual pairs are appearing for a short lifetime. Appearance and disappearance of fermion couples have to obey the observed Lorentz invariance of space-time. Formation of condensates which are Lorentz scalars with vanishing charge is allowed. Thus fermion condensates must be of the form $\langle\overline\Psi\Psi\rangle$ where $\Psi$ defines the fermion field. Similarly, a tensor field can only have a scalar expectation value.

All the virtual particles can exist only inside a Lorentz space-time shell defined by the Heisenberg uncertainty principle: $\Delta x \Delta p \ge \frac{\hbar}{2}$ and $\Delta E \Delta t  \ge \frac{\hbar}{2}$.
The pairs are CP-symmetric, which allow them to keep the total angular momentum, colour and spin zero values. Considering the Pauli exclusion principle, the virtual fermions in different pairs with the same type and spin cannot overlap, therefore the ensemble of virtual particle pairs tends to expand, like a gas in an infinite vacuum. We expect that the virtual pair gas is in thermal equilibrium. The energy exchange between the virtual pairs can be possible due to, for example, the presence of the real and virtual photons and their continuous absorption and reemission by virtual pairs. Also some indirect interaction between the virtual pairs could be possible thanks to the Pauli exclusion principle itself. Pairs can not appear in the place already occupied by other pairs and thus the limited free space can be only occupied by the particles with a particular energy/momentum since they are connected with the Heisenberg uncertainty. We expect that the virtual gas expansion may drive the Universe expansion and this process is infinite, going with a decreasing acceleration and cooling temperature.

In this model, as an extension of the commonly accepted Standard Model, the real photons are excitations of virtual particles, realised as harmonic oscillations. In this paper it is shown how, from this definition, we can describe the particle-wave duality and arrive to Quantum Field Theory (QFT) wave-function (Section~\ref{sec:WavePhoton}). The vacuum permeability and permittivity can be explained from the gas of virtual particles and its statistical properties variation in the presence of the external fields (Section~\ref{sec:MuEpsi}). The last part describes the consequences on the real photon propagation, and the measurements that can be performed to test this model.

\section{Wave Function and Photon Propagation}
\label{sec:WavePhoton}
The excitation of a virtual pair can appear as oscillations of fermions. This definition allows introducing the Hamiltonian of a harmonic oscillator for virtual pair description. A propagating real photon transmits its full energy and momentum to the virtual pairs creating one dimension oscillators with a frequency corresponding to the photon energy $E=h\nu$. We assume that during the real photon propagation between the pairs, the oscillation phase is transferred as well. From this point, considering a continuous space approximation this model allows the development of the classical quantum and quantum field theories using the Schr\"odinger equation to find the photon wave function~\cite{RefQFT}.

From this model it naturally appears that the real photon properties cannot be measured below the Heisenberg uncertainties of the excited virtual pairs. We can already notice some other important properties:
\begin{itemize}
\item while the virtual particle properties are still in debate~\cite{RefVirtPart}, in this model we consider that virtual particles conserve the energy and momentum and have the same properties as the real ones, which may include the mass. In consequence, the only difference with real particles is their short lifetime defined by the Heisenberg uncertainty;
\item being associated to fermion pairs, the propagation of a real photon can be described as leaps of distances with time delays arising from virtual pairs size and lifetime;
\item at each photon propagation leap the virtual pair annihilation can be considered as a secondary emission point with a decreasing density, which corresponds to Huygens-Fresnel wave principle;
\item by entanglement, locally, each oscillation contains all the information about the propagating real photon;
\item the oscillator can be defined as a discrete wave at each position $\vec r$ at a time $t$. The oscillator amplitude $\Psi (\vec r, t)$ corresponds to the probability of observing the real particle in the volume, $(\Delta x)^3$, occupied by oscillating virtual pair as  $\Delta \mathscr{P}(\vec r, t)=C|\Psi (\vec r, t)|^2 (\Delta x)^3$, where C is a normalisation constant;
\item the discrete function $\Psi (\vec r, t)$ constructed from the oscillator amplitudes of virtual pairs may be approximated as continuous function on the scales above the Heisenberg uncertainty and it contains the full information about the real particle, analogously to wave function in QFT.
\end{itemize}

\section{The Stochastic Vacuum Description}
\label{sec:StochasticVacuum}
Additionally to the commonly known virtual photons, the vacuum is filled with virtual fermions that form the propagation medium for the real photons. In total, there are 21 charged fermion pair species, noted $i$, considering the charged lepton families and the quark families with their colour states.

Each fermion in a virtual pair is defined by the size, $x_i$, and no other fermion of the same family and spin state can occupy the volume of $x_i^3$, following the Pauli exclusion principle. We assume that the virtual pair energy, $E$, and its lifetime, $T$, are connected thanks to the Heisenberg uncertainty in the following way:
\begin{equation}
ET=h.
\label{eq:HeisET}
\end{equation}
Each virtual fermion momentum, $p$, and size, $x$, are also connected in the similar way:
\begin{equation}
px=h.
\label{eq:Heispx}
\end{equation}
Moreover, both virtual fermions in a pair have equal energies, $\varepsilon$, and momentum absolute values, so $E=2\varepsilon$. If  energy momentum relation holds for virtual particles, then, for each fermion the energy can be expressed through the momentum:
\begin{equation}
\varepsilon^2=(mc^2)^2+(pc)^2.
\end{equation}
In this expression $m$ is the mass of a virtual fermion that can be equal to the mass of a corresponding real fermion. It will be shown later, however, that some properties are better explained with zero virtual fermion masses.

The scalar nature of the Higgs field may allow the formation of fermions condensates $\langle\overline\Psi\Psi\rangle$, so it can be the origin of the virtual pairs. The average expected value in the vacuum of this field is given by $\upsilon = 1/\sqrt{\sqrt 2 G_F^0} \approx 246.22\, \rm{GeV}$, where $G_F^0$ is the reduced Fermi constant. We speculate that this energy corresponds to the temperature of the virtual pair gas being in thermal equilibrium, $kT_\nu=\upsilon$. 

In order to obtain the virtual fermion density distribution, one can use grand thermodynamic potential as following:
\begin{equation}
\Omega=-PV,
\end{equation}
\begin{equation}
n= \frac{\partial P}{\partial \mu} = - \frac{1}{V} \frac{\partial \Omega}{\partial \mu}.
\label{eq:n:Gibbs}
\end{equation}
In the latter expression $\mu$ is the chemical potential and the pressure, $P$, is composed of fermion and antifermion pressures as following~\cite[Equation (18)]{Srivastava}:
\begin{multline}
P=P_++P_-=kT_\nu\left[\int_0^{+\infty} \frac{4\pi p^2 dp}{h^3} ( \ln(1+e^{(\mu-\varepsilon)/kT_\nu}) + \right. \\
\left.\ln(1+e^{(-\mu-\varepsilon)/kT_\nu}))\right].
\label{eq:pDirac}
\end{multline}
The pressure of fermions, $P_+$, and the pressure of antifermions, $P_-$ have the same expressions apart the chemical potential that has the same value but the opposite sign due to the production in couples fermion-antifermion. Note, that the original formula~\cite[Equation (18)]{Srivastava} contains a factor 2 since each energy level can be occupied with two different spin states of the fermion and antifermion. However, since the virtual particles are always produced in pairs that should annihilate with a zero value of the total spin, only pairs with antiparallel spins for particles and antiparticles are allowed, so we believe this reduces the number of the states by a factor of two.

From Equations (\ref{eq:n:Gibbs},\ref{eq:pDirac}) the density becomes: 
\begin{equation}
n= \int_0^{+\infty} \frac{4\pi p^2 dp}{h^3}\left(\frac{1}{e^{(\varepsilon-\mu)/kT_\nu}+1}-\frac{1}{e^{(\varepsilon+\mu)/kT_\nu}+1}\right).
\label{eq:lepton_densitiy}
\end{equation}
Note that this density is negative for antiparticles, so is does not describe the virtual particle density, but rather the lepton number density. Moreover, in order to guarantee that the lepton density is zero, $\mu=0$ should be used in order to equalise the number of particles and antiparticles for each energy. Finally, the density of virtual particles as a function of momentum can be written taking one of the exponential terms in Equation~(\ref{eq:lepton_densitiy}) and assuming $\mu=0$:
\begin{equation}
n_\textrm{pair}(p)= \frac{4\pi p^2}{h^3}\frac{1}{e^{\varepsilon/kT_\nu}+1}.
\label{eq:pair_densitiy}
\end{equation}
In this expression $\varepsilon$, $p$ are the energy and the momentum of fermions, contrary to the values corresponding to the virtual pairs. To account for different fermion families all the equation sin this section should be summed over 21 fermion species.

\section{The vacuum permeability and permittivity}
\label{sec:MuEpsi}
It is experimentally known that the vacuum has a non-zero value of permeability,  $\mu_0$, and permittivity, $\epsilon_0$.
As proposed in \cite{RefUrban}, the  permeability and permittivity can be fundamental properties of the stochastic vacuum.

Being composed by opposed charges and spins each virtual pair carries doubled spin magnetic moments related to each particle, $\beta$, and the dipole moment, $\omega$. In the presence of the magnetic (electric) field, the statistical distribution of virtual pairs and fermions/antifermions is altered and this results in a non-null magnetic moment density, $\vec M$, (polarisation, $\vec P$). Note that, following discussions in~\cite{RefUrban}, $\vec M$ and $\vec P$ do not have a commonly accepted meaning of a magnetic density of medium and material polarisation because they are considered to be null in vacuum. Instead their non-zero values are ascribed to a vacuum behaving like a gas of virtual pairs. The vacuum permeability and permittivity can be derived from those values using the following expressions: $\vec M= 1/\mu_0 \vec B$ and $\vec P = \epsilon_0 \vec E$.

Following~\cite[Equations (59.1), (59.4),  (59.12)]{LL}:
\begin{align}
|\vec M| = -(\partial\Omega/\partial B)_{T,V,\mu}, \\
1/\mu_0=-\frac{2}{3} \frac{\beta^2}{V} \frac{\partial^2 \Omega_0}{\partial \mu}=\frac{2}{3} \beta^2 \frac{\partial n}{\partial \mu},
\label{eq:mu0_Gibbs}
\end{align}
where $\Omega_0$ is the grand potential in the absence of the external fields and $n$ can be taken from Equation~(\ref{eq:lepton_densitiy}).

The relativistic energy for an electron in the magnetic field can be expressed as following~\cite{RelMM}:
\begin{multline}
E_M=\sqrt{m^2c^4+p_z^2c^2 + 2e\hbar cB(n+1/2 -gs/2)} \\
\approx E + \frac{e\hbar c B}{2E} (2n+1-gs),
\label{eq:relenergyB}
\end{multline}
where we take the first order approximation valid for common fields $\beta_B B \ll m_ec^2$ ($\beta_B$ is the Bohr magneton).
This approximation can be compared to the non-relativistic energy levels used in~\cite{LL} to derive Equation~(\ref{eq:mu0_Gibbs}):
\begin{equation}
E_M^{NR}=\frac{p_z^2}{2m}+\frac{e\hbar B}{2mc}(2n+1-gs).
\end{equation}

Comparing relativistic and non-relativistic expressions one can see that the following relativistic magnetic moment can be used to match them:
\begin{equation}
\beta=\frac{e\hbar}{2\varepsilon/c},
\label{eq:relMM}
\end{equation}
where respect to the Bohr magneton, $\beta_B = e\hbar/(2mc)$, $mc$ is substituted with $\varepsilon/c$. The evaluations in~\cite{LL} should remain valid even for energy dependent $\beta$. However, the integration in $\Omega$ or $N$ over the energy will now include $\beta^2$ as well\footnote{We were able to check the validity of math operations to derive the paramagnetic part. For the diamagnetic part, the necessary condition $\beta |\vec B| \ll kT$ is satisfied too. However, we cannot guarantee the validity of the complete evaluation in the case where $\beta$ is in function of the energy.}. Equation (\ref{eq:mu0_Gibbs}) then becomes:
\begin{multline}
1/\mu_0=\frac{2}{3} \int_0^{+\infty} \beta(\varepsilon)^2 \left.\frac{\partial n(p)}{\partial \mu} \right|_{\mu=0} dp= \\
\sum_i \frac{2}{3} \int_0^{+\infty} \beta(\varepsilon)^2 \frac{2\cdot4\pi p^2 dp}{h^3} \frac{e^{\varepsilon/kT_\nu}}{e^{\varepsilon/kT_\nu}+1}
\label{eq:mu0_final}
\end{multline}
The sum over different virtual pair types, $i$, is added since the overall vacuum magnetisation $|\vec M|$ is linearly composed from the densities of magnetic dipole moments of different types.

In classical physics the electric and magnetic dipole potential energies have the same expressions: $U= -\vec\omega\cdot\vec E$ and $U=-\vec\beta\cdot\vec B$, so we expect the energy levels and the state function to change in the presence of the magnetic and electric fields in the same manner. This brings for the vacuum permittivity the following expression:
\begin{equation}
\epsilon_0=\frac{2}{3} \int_0^{+\infty} (\omega/2)^2 \left.\frac{\partial n(p)}{\partial \mu} \right|_{\mu=0} dp.
\label{eq:epsi0_Gibbs}
\end{equation}

Note that we substituted  $\beta$ with $\omega/2$ since the magnetic moment was related to a fermion or an antifermion while the dipole moment is related to a whole couple. 

From Maxwell’s equations, the connection between the measured speed of light waves, $c$, and vacuum properties is the following:
\begin{equation}
c=\frac{1}{\sqrt{\mu_0\epsilon_0}}=\sqrt{\frac{1/\mu_0}{\epsilon_0}}.
\end{equation}

This equation together with Equations~(\ref{eq:mu0_Gibbs},\ref{eq:epsi0_Gibbs}) can be satisfied if $\beta=(\omega/2) c$. To define the dipole moment one can imagine a virtual pair being composed of two charges separated by the average distance $x$:
\begin{equation} 
\omega = Q ex= Qe\frac{h}{p}.
\label{eq:relDM}
\end{equation}

Actually if the virtual fermions were randomly distributed in the filled sphere of radius $x$, the average distance would be $36/35x$. The sphere approximation, however, does not fit with $x^3$ volume assumed in the statistical distributions.

Comparing this expression with Equation~(\ref{eq:relMM}) one can see similarities considering for electrons $Q_i=1$ and assuming $\varepsilon \approx pc$ which can become exact if virtual fermions have zero masses.  In the magnetic moment expression, Equation (\ref{eq:relMM}), however, $\hbar$ is different from $h$ in the dipole moment expression, Equation~(\ref{eq:relDM}). This can be unified to $\hbar$ if the original relation between the virtual fermion momentum and its size is rewritten as $px = \hbar$ or the factor $2\pi$ is actually arriving from the difference of the fermion size and the average distance between fermion and antifermion. Note, that it is hard to choose the right expressions {\it a priori} due to the ambiguity for relativistic magnetic moment and dipole interactions as well as their Lorenz transformations~\cite{RelMM,pmtrans}.

From Equation~(\ref{eq:mu0_final}) assuming $\beta_i=Q_ieh/(2\varepsilon/c)$ we obtain a value of $\mu_0\approx5.6\cdot 10^{-6}$\,H/m. This value is 4.4 times bigger than the measured magnetic constant value. In our opinion a non-zero vacuum magnetisation is already an interesting feature of this theory. In fact, the model has no free parameters and the fact that the $\mu_0$ value results of the same order as the measured one is very intriguing. Interestingly, assuming zero fermion masses in virtual couples makes the difference $4.1$. The remaining difference may be ascribed due to approximations used to derive Equations~(\ref{eq:mu0_Gibbs}, ~\ref{eq:relenergyB}). This can also hint for missing non standard or sterile fermions, which could increase the vacuum magnetisation. However, this may contradict fermion loop corrections in quantum field theory. In particular, the factor close to four can be explained by the presence of two more quantum numbers (positive, negative) distinguishing standard fermions from the hidden ones with all the remaining properties being the same. This would make $\mu_0$ only $2\%$ different from the measured value assuming zero fermion masses.  

Notice, however, that using more natural expression for $\beta$ from Equation~(\ref{eq:relMM}) with $\hbar$ would result in a 180 times smaller vacuum magnetisation than currently observed. If one also uses $\hbar$ in Equation (\ref{eq:pDirac}) the magnetic moment would be approximately $0.8\cdot10^{-6}$\,H/m which is 0.7 smaller than the measured value. Such substitution, however, would violate Planck's law and the photon energy dependence from its frequency. Using the non-relativistic magnetic moment, $\beta_i=Q_ieh/(2m_ic)$, results in ten orders of magnitude higher vacuum magnetisation. Finally, as a side remark, our trials to calculate vacuum polarisation (magnetisation) as an integral of electric (magnetic) dipole moment projection on the field axis and multiplied by the fermion/pair density were not successful. No analytical solution was obtained and for numerical integration the precision was lost due to oscillating terms. This problem is known already for a non-relativistic case~\cite{complete_magnetisation}.

\section{Photon Propagation and Propagation Time Dispersion}
The photon propagation is composed from subsequent perturbations of charged virtual fermion pairs. Since the space is filled with virtual particles at a high density, and since the family and spin states allow overlapping of pairs, there is no expected additional time or free fly of the photon in absolute vacuum like in traditional physics or in the theory described in~\cite{RefUrban}.  
Assuming the random time of the pair excitation the average time during which the pair was excited and annihilated is $\frac{T}{2}=\frac{h}{2E}=\frac{h}{4\varepsilon}$. In order to conserve the light propagation velocity as $c$ the propagation distance should be $\frac{x}{4}$ since then the  velocity becomes $v=\frac{\varepsilon}{p}$. The latter ratio is close to $c$ for the relativistic pairs ($E \gg 2mc^2$) or it becomes equal to $c$ if virtual fermions have zero masses. 

If one assumes that the energy of each fermion is increased by $\varepsilon_\gamma/2$ then the average speed can be obtained as:
\begin{equation}
v = \frac{\int_0^{+\infty} \frac{c}{\sqrt{(\varepsilon+\varepsilon_\gamma/2)^2-(mc^2)^2}} n_\textrm{pair}(p) dp}{\int_0^{+\infty} \frac{1}{\varepsilon+\varepsilon_\gamma/2} n_\textrm{pair}(p) dp}.
\end{equation}
For $\varepsilon_\gamma=0$ the difference $(v-c)/c$ is about $4\cdot10^{-7}$, decreasing down to one at $\varepsilon_\gamma\to\infty$. In this model it is expected that for a signal composed from photons with $\varepsilon_\gamma=[30,200]$~keV the photon speed variation would be on the order of $\Delta v/c\approx1.4\cdot10^{-11}$ which is well above the existing limits obtained from the Gamma Ray Burst (GRB) 980703 observation, $\Delta v/c < 6.3\cdot10^{-21}$~\cite{GRB980703}. The recent GRB data compilation~\cite{GRB2016} supports the Lorentz invariance violation and suggests a linear form of light speed variation with dependence $v=c(1-\varepsilon_\gamma/E_{LV}$) with $E_{LV}=3.6\cdot10^{17}$\,GeV. Such dependency also has a decreasing trend with growing photon energy. However, the analysed high energy GRB photons in the range $\varepsilon_\gamma=[1,50]$~GeV would have $\Delta v/c\approx1.3\cdot10^{-16}$ while in our model it is $\Delta v/c\approx1.5\cdot10^{-7}$ for the same photon energy range, which is a well-larger variation. The inconsistencies with the GRB observations can be avoided in several ways. 
\begin{itemize}
\item One could doubt the same VEV and the vacuum temperature at earlier Universe correspondent to the time of the GRB explosions and the light propagation. 
\item It could be proposed that the photons do not propagate following a straight line but rather following excitations of virtual pairs which are not perfectly aligned. If the $\epsilon_\gamma$ does not alternate the size and the lifetime of the virtual particles, the photons velocity is constant. This can be valid since the Heisenberg uncertainties could be valid only for the excess of the energy/momentum respect to the ``real'' values. Alternatively, since the sizes of the virtual particles are modified, the trajectory could become straighter and compensate for the smaller velocity.
\item If the virtual fermions have lower masses compared to the real fermions, the velocity variation would be decreased reaching $c$ value independent from $\varepsilon_\gamma$ for zero fermion masses.
\end{itemize}

The stochastic dispersion of the photon propagation time is one of the most significant and measurable effects distinguishing this theory from the QFT. This dispersion depends on the number of encountered virtual pairs along the path and the variation of the delay at each step.

The pair type excited at each photon step is defined by the probability:
\begin{equation}
P_i(p)=\frac{n_i(p)}{\sum_j \int_0^{+\infty} n_j(p) dp},
\end{equation}
where $n_k(p)$ is the virtual pair density for each family, defined in Equation~(\ref{eq:pair_densitiy}). Ignoring the additional energy $\varepsilon_\gamma/2$, the total length can be composed of $x/4=h/(4p)$ steps in the following way:
\begin{equation}
    L=N \frac{\sum_{i} \int_{0}^{+\infty} n_i(p) \frac{h}{4p} dp}{\sum_{j} \int_{0}^{+\infty} n_j(p) dp},
    \label{eq:totpath}
\end{equation}
from where the average number of steps can be recovered:
\begin{equation}
    N = \frac{4L}{h} \frac{\sum_{j} \int_{0}^{+\infty} n_j(p) dp}{\sum_{i} \int_{0}^{+\infty} n_i(p) \frac{1}{p} dp}.
\end{equation}

Assuming the fact that the photon excites pairs that live in average $T/2$ at random time and the lifetime distribution is flat, the fluctuation of the photon delay at each step is:
\begin{multline}
\sigma_i(p) = \frac{T}{2\sqrt{3}}=\frac{h}{2\sqrt{3}E}=\\
\frac{h}{4\sqrt{3}\varepsilon}=\frac{h}{4\sqrt{3((m_ic^2)^2+(pc)^2)}}.
\end{multline}

Thus the total dispersion:
\begin{equation}
\begin{aligned}
\sigma = \sqrt{N \sum_i \int_{0}^{+\infty} n_i(p) \sigma_i^2(p) dp}=\\
\sqrt{L\frac{h}{12}\frac{\sum_i \int_{0}^{+\infty} n_i(p) \frac{1}{(m_ic^2)^2+(pc)^2} dp}{\sum_{j} \int_{0}^{+\infty} n_j(p) \frac{1}{p} dp} }\approx 1.9\,\mathrm{as\,m}^{-1/2}.
\end{aligned}
\end{equation}

Note that if one uses $\hbar$ instead of $h$ in the relations~(\ref{eq:HeisET},\ref{eq:Heispx}) this dispersion decreases by a factor of $\sqrt{1/(2\pi)}$ and becomes approximately $0.8\,\mathrm{as\, m}^{-1/2}$.

\section{Experimental Observation of the Stochastic Vacuum}
The observation of stochastic photon propagation time dispersion would consolidate the model with a concept of photon propagation described as leaps between virtual fermion pairs. For the model in~\cite{RefUrban} the expected fluctuation is more than an order of magnitude higher compared to the fluctuation evaluated here. Both models provide values that can be measured with the available technologies, oppositely to fluctuations expected in other theories~\cite{IntrinsicLimit}. 

Nowadays, the strongest constraints are established by astrophysical observations, mainly GRBs and pulsars~\cite{Urban8,Urban9}. The current limits are at $0.2-0.3\,\mathrm{fs\,m}^{-1/2}$.

In~\cite{RefUrban} and in the presented model, the time resolution has a stronger impact than the photon path length, which favours measurements in laboratory. 
In fact, astrophysical measurements are based on the events observation with a duration of the order of $10^{-3}\,\mathrm{s}$ at distances of megaparsecs (a factor $10^{22/2}=10^{11}\,\mathrm{m}^{1/2}$). The current state of art of laser technologies allows the generation of femtosecond light pulses and a measurement precision is at the order of $10^{-15}\,\mathrm{s}$, on a distance of tens of kilometres (a factor $10^{4/2}=10^{2}\,\mathrm{m}^{1/2}$).
As a result, the astrophysical measurement precision is at the order of $10^{-14}\mathrm{fs\,m}^{-1/2}$, while the laboratory measurement precision can reach $10^{-17}\mathrm{fs\,m}^{-1/2}$.

The experiment able to reach the required sensitivity can be realised with femto-second laser pulses propagating in a multi-pass vacuum cavity, which can have length of several kilometres (as used by the Virgo/LIGO experiments~\cite{Virgo,LIGO}). Under these conditions the order of magnitude of the arrival time dispersion predicted by this theory can reach femtoseconds (e.g. $\sigma \approx 1$\,fs for a 4\,km cavity with  70 reflections). The pulse width characterisation is possible thanks to an auto-correlation measurement system such as FROG~\cite{frogcite}. This kind of system allows measuring the pulse width for different wavelengths, which can provide information on systematic effects since the expected velocity variation with wavelength is negligible for such measurements.

\section{Conclusions and Discussions}
The Heisenberg uncertainty principle allows virtual pairs with energy $E$ to appear for times $\ge\hbar/E$ in the vacuum. Starting from stochastic vacuum model of the space filled with virtual particles, we hypothesised that real photons propagation can be explained as a continuous absorption by virtual fermion pairs and emission after the pairs annihilation. Wave-like properties appearing in this model follow Huygens-Fresnel principle since every annihilating pair becomes an emission point.

At the first glance, the virtual pairs should appear and annihilate independently in the absence of real particles. However, it is natural to propose that the virtual fermions should obey Pauli exclusion principle similar to the real particle and this already makes an interconnection of the pair production in a particular space from the nearby pairs. As a consequence we hypothesise that the virtual pairs as well as the virtual photons are in thermal equilibrium. Additionally, this drives the expansion of the virtual particles gas, which could explain the Universe expansion. The photons distribution should follow Planck's law while the virtual pair energies are distributed according to the equation of state of a Fermi gas. For the temperature we propose the vacuum expectation value of the Higgs field. Assuming this, we formally introduce a new concept of the {\it vacuum temperature}. In classical physics the vacuum has undefined temperature since no carriers are present.

Going further one could imagine that real photons exist only at the production and the interaction points while during the propagation information about the photons is spread by the virtual pair excitation. The simplest excitation of the virtual pair is a harmonic oscillation and it corresponds to the $U(1)$ group, which is enough to explain the photon properties. For other real particles, the theory can be extended and the properties such as {\it SU}(2) for the spin may correspond to a rotation of the pairs and so on.

Vacuum permittivity and permeability appear naturally assuming every virtual fermion carries a magnetic moment and each pair has a dipole moment. The obtained value of magnetic constant, $\mu_0$, is about 4.5 times higher than the measured value. The remaining difference may suggest for missing fermions or it can be due to the approximations used. We believe that the same order of magnitude is still a success of this theory that has no parameters to tune apart from $h/\hbar$ choice in the Heisenberg uncertainty, temperature and the relativistic magnetic moment definition. The values of the fermion magnetic moment, $\beta$, and the virtual pair dipole moment, $\omega$, are connected as $\beta=(\omega/2)c$ to guarantee the speed of light value calculated from $\mu_0$ and $\epsilon_0$. This connection can be retrieved if those moments are written from the first principles and the assumption that the virtual pair size and virtual fermion momentum are connected as $px=h$ appearing from the Heisenberg uncertainty.

In this theory it appears that photon propagation speed or travel time to cover some distance may fluctuate since it depends on the number of virtual pairs involved in the propagation. The effect can be measurable on the kilometre scale distances with femtosecond pulse lasers. Phase velocity remains, instead, more stable~\cite{answer} which is in agreement with current interferometric measurements. It is interesting to note that nowadays there are existing kilometre scale cavities but they are used with ultra-stable continuous lasers (Virgo/LIGO experiments~\cite{Virgo,LIGO}), while the ultra-fast lasers are usually used at very short range (meter scale).

The theory presented here has several differences respect to the theory in~\cite{RefUrban}. These are mainly the following: 
\begin{itemize}
\item the virtual pairs energy follows the equation of state of a Fermi gas with no free parameters instead of tuned fixed virtual pair energies or tuned $dE/E^2$ spectrum,
\item the abundances of fermion families follows the equation of state instead of abundances driven by the maximum densities,
\item the calculation of the vacuum permeability through the equation of state alternation in the presence of the magnetic field  instead of the average magnetic moment calculation with a modification of a lifetime due to the presence of the magnetic field;
\item for the photon propagation no assumption of the cross-section is needed; it is assumed that photons make leaps from an annihilated pair immediately to the next one with no free path.
\end{itemize}.

As a result of the first two changes, the obtained fluctuation expectations are on the order of $1.9$~as\,m$^{-1/2}$  which is more than an order of magnitude below respect to the predictions in~\cite{RefUrban}. The values predicted here are still within reach of the currently available technologies. Considering, for example, that for the path in the tunnel of LIGO with 70 reflections and the commercially available lasers with an initial pulse of 4~fs (FWHM) the expected pulse broadening up to 4.6~fs (FWHM) would be measurable with the FROG techniques. 

It is commonly accepted that the virtual fermion masses do not correspond to the real fermion masses and the virtual particles are thus called off-mass-shell. The results obtained in the framework of this theory looks slightly more appealing if the zero virtual fermion masses are assumed. In particular, the photon propagation velocity becomes exact as $c$; $\mu_0$ obtained in the framework of this theory becomes slightly closer to the measured value; speed of light from $\mu_0$ and $\epsilon_0$ becomes also exact as $c$ (the equation for this velocity is different from the photon propagation velocity here). There are, however, no unexplainable contradictions with the existing observations if the mass of the virtual fermions are kept equal to the corresponding real particle masses. Note, that same advantages can be obtained if instead of assuming zero fermion masses, one substitutes $px=h$ to $(\varepsilon/c)x=h$.

Further studies of the virtual pair gas dynamics could be promising to explain different Cosmological and Quantum Field properties such as the Universe expansion and gravity. We hope that the provided observations and offered possibilities for both theory and experiments are intriguing and will lead to more results in this field.




\end{document}